\documentclass[prx,sd,amsmath,amssymb,reprint]{revtex4-1}

\usepackage{graphicx}
\usepackage{bm}
\usepackage{color}

\begin{document}


\title{Magic numbers for vibrational frequency of charged particles on a sphere}

\author{Shota Ono}
\email{shota\_o@gifu-u.ac.jp}
\affiliation{Department of Electrical, Electronic and Computer Engineering, Gifu University, Gifu 501-1193, Japan}

\begin{abstract}
Finding minimum energy distribution of $N$ charges on a sphere is known as the Thomson problem. Here, we study the vibrational properties of the $N$ charges in the lowest energy state within the harmonic approximation for $10\le N\le 200$ and for selected sizes up to $N=372$. The maximum frequency $\omega_{\rm max}$ increases with $N^{3/4}$, which is rationalized by studying the lattice dynamics of a two-dimensional triangular lattice. The $N$-dependence of $\omega_{\rm max}$ identifies magic numbers of $N=12, 32, 72, 132, 192, 212, 272, 282$, and 372, reflecting both a strong degeneracy of one-particle energies and an icosahedral structure that the $N$ charges form. $N=122$ is not identified as a magic number for $\omega_{\rm max}$ because the former condition is not satisfied. The magic number concept can hold even when an average of high frequencies is considered. The maximum frequency mode at the magic numbers has no anomalously large oscillation amplitude (i.e., not a defect mode).
\end{abstract}

\maketitle

\section{Introduction}
Magic numbers play an important role in understanding the energetic stability of nanoparticles or clusters. The structures at magic numbers are highly symmetric and have particularly low energies, as shown in unary and binary Lennard-Jones clusters \cite{romero,doye}, fullerenes \cite{zhang}, and other metallic clusters; see Ref.~\cite{NP} for an extensive review on nanoclusters. More recently, the magic number concept has been extended to colloidal clusters, irrespective to negligibly small inter-particle interactions \cite{wang}. 

In general, the ground-state structure for a cluster is equal to the global minimum of potential energy surface (PES) characterized by $Nd$ dimensional space, where $N$ is the number of particles and $d$ is a space dimension in which the particles can move. At finite temperature, the clusters will oscillate around the minimum of the PES. Such a vibrational property will be sensitive to the geometry and/or curvature around the minimum of the PES because the force constant matrix is constructed by the Hessian of the potential energy as a function of particle positions. Magic numbers will thus appear in the sequence of vibrational frequency as a function of $N$. However, to the best of our knowledge, the presence of such anomalies has not been explored. 

As a model, we consider $N$ charges on a unit sphere that is a platform in the Thomson problem and/or the Smale's seventh problem \cite{smale}: to determine the minimum energy configuration of $N$ charges confined to the surface of a unit sphere. The global and local minima of the PES have been investigated for this system \cite{morris,alt,alt2,wales,wales2009,mehta}, while many related problems have also been studied \cite{peters1,peters2,peters3,yang,robinson,batle,yangQM,gnidovec}. Through the total energy calculations for $N\le 200$, $N=12, 32, 72, 122, 132, 137, 146, 182, 187$, and 192 have been identified to be magic numbers \cite{morris}. 

In this paper, we demonstrate that magic numbers are present in the vibrational frequency of $N$ charges on a unit sphere. By considering the range $10\le N\le 200$ and studying the lattice vibration of two-dimensional (2D) triangular lattice, we find that the maximum frequency is significantly small at $N=12, 32, 72, 132$, and $192$. On the other hand, no anomalies are found at $N=122, 137, 146, 182$, and 187 that are identified as the magic numbers from total energy calculations. We also identify $N=212, 272, 282$, and $372$ as magic numbers by calculating the vibrational frequency. The presence of magic numbers for the maximum frequency can be attributed to a strong degeneracy of the one-particle energies that will be derived from its icosahedral structures of distribution of charges. The magic number concept is valid as long as an average of high frequencies is considered. Finally, the maximum frequency mode at the magic numbers can be assigned to a delocalized mode.

\section{Theory}
\label{sec:theory}
\subsection{Lattice dynamics on a sphere}
We apply the theory of lattice dynamics (see Ref.~\cite{ziman} for example) to vibrations of $N$ charged particles confined to the surface of a sphere with a unit radius. The Lagrangian of the system is given by
\begin{eqnarray}
 L = \frac{1}{2} \sum_{i=1}^{N} \left[\left( \frac{d\theta_i}{dt}\right)^2 
 + \sin^2\theta_i \left( \frac{d\phi_i}{dt}\right)^2 \right] - V,
\label{eq:Lag}
\end{eqnarray} 
where $\theta_i$ and $\phi_i$ are the spherical coordinates of charge $i$ with a unit mass. The first term is the kinetic energy and the second term $V$ is the potential energy given by
\begin{eqnarray}
 V = \sum_{i=1}^{N} \varepsilon_i
\label{eq:Etot}
\end{eqnarray}
with the one-particle energy
\begin{eqnarray}
 \varepsilon_i = \frac{1}{2} \sum_{j\ne i}^{N} \frac{1}{r_{ij}}.
 \label{eq:one}
\end{eqnarray}
The factor of $1/2$ in Eq.~(\ref{eq:one}) accounts for the double counting of the interaction energy between charges $i$ and $j$ that is inversely proportional to the Euclidean distance $r_{ij}$ expressed by
\begin{eqnarray}
 r_{ij} = \sqrt{2} \left[ 1 - \sin \theta_i \sin \theta_j \cos \Delta\phi_{ij} 
 - \cos \theta_i \cos\theta_j \right]^{\frac{1}{2}}
\label{eq:rij}
\end{eqnarray}
with $\Delta\phi_{ij} = \phi_i - \phi_j$. With the analytical mechanics, the equations of motion for $\theta_i$ and $\phi_i$ are given by
\begin{eqnarray}
& & \frac{d^2 \theta_i}{dt^2}  
 = - \frac{\partial V}{\partial \theta_i} + \sin\theta_i \cos\theta_i \left( \frac{d\phi_i}{dt}\right)^2,
 \label{eq:ddtheta}
\\
& &  \sin^2\theta_i \frac{d^2 \phi_i}{dt^2}
+ 2  \sin\theta_i \cos\theta_i \frac{d\theta_i}{dt} \frac{d\phi_i}{dt}
 = - \frac{\partial V}{\partial \phi_i}.
 \label{eq:ddphi}
\end{eqnarray}
Using Eq.~(\ref{eq:ddphi}), we can show the relation
\begin{eqnarray}
\frac{d}{dt} \left( \sum_i \sin^2\theta_i \frac{d \phi_i}{dt} \right)
 = - \sum_i \frac{\partial V}{\partial \phi_i} = 0,
\end{eqnarray}
which states that the total angular momentum is conserved. The coupled Eqs.~(\ref{eq:ddtheta}) and (\ref{eq:ddphi}) determine the time ($t$) evolution of $N$ charges on a sphere given an initial condition. 

In this paper, we study the dynamics of the $N$ charges within the harmonic approximation around the equilibrium configurations $(\theta_{i}^{0},\phi_{i}^{0})$ with $i=1,\cdots,N$. By assuming $\theta_i = \theta_{i}^{0} + u_{i\theta}$ and $\phi_i = \phi_{i}^{0} + u_{i\phi}$ with the displacements $(u_{i\theta},u_{i\phi})$, the equations of motion can be written as
\begin{eqnarray}
m_{i\alpha} \frac{d^2 u_{i\alpha}}{dt^2}  
 = - \sum_{j\beta} D_{\alpha\beta}^{ij} u_{j\beta},
\end{eqnarray}
where the terms proportional to $(d\phi_i/dt)^2$ and $(d\theta_i/dt)(d\phi_i/dt)$ in Eqs.~(\ref{eq:ddtheta}) and (\ref{eq:ddphi}) are omitted. $\alpha (\beta)$ indicates $\theta$ or $\phi$, and $m_{i\alpha}$ is an effective mass defined as $m_{i\alpha} = \delta_{\alpha\theta} + \delta_{\alpha\phi}\sin^2\theta_{i}^{0}$ that originates from the confinement of the surface of the sphere. The force constant matrix is given by 
\begin{eqnarray}
D_{\alpha\beta}^{ij} = D_{\beta\alpha}^{ji} =
 \frac{\partial^2 V}{\partial \alpha_i \partial \beta_j} \Big\vert_0,
 \label{eq:dynmat}
\end{eqnarray}
where the derivative is taken at the equilibrium configurations. Assuming a stationary solution $u_{i\alpha}(t) = \epsilon_{i\alpha} e^{i\omega t}$ with the frequency $\omega$ and the polarization $\epsilon_{i\alpha}$, one obtains the eigenvalue equation
\begin{eqnarray}
m_{i\alpha} \omega^2 \epsilon_{i\alpha} 
&=& \sum_{j\beta}
D_{\alpha\beta}^{ij} \epsilon_{j\beta}.
\label{eq:eigen}
\end{eqnarray}

Analytical expressions of Eq.~(\ref{eq:dynmat}) are given by
\begin{widetext}
\begin{eqnarray}
 D_{\theta\theta}^{ii} 
 &=& \sum_{j(\ne i)}
 \left( 
 \frac{3}{r_{ij}^{5}} 
 \left[ - \cos\theta_i \sin \theta_j \cos\Delta\phi_{ij} + \sin\theta_i \cos\theta_j \right]^2
 - \frac{1}{r_{ij}^{3}} 
 \left[ \sin\theta_i \sin \theta_j \cos\Delta\phi_{ij} + \cos\theta_i \cos\theta_j \right]
 \right)
 \label{eq:Diitt}
 \\
 D_{\theta\phi}^{ii} 
 &=& \sum_{j(\ne i)}
 \left(
 \frac{3}{r_{ij}^{5}} 
 \left[ - \cos\theta_i \sin \theta_j \cos\Delta\phi_{ij} + \sin\theta_i \cos\theta_j \right]
 \sin\theta_i \sin \theta_j \sin\Delta\phi_{ij} 
 - \frac{1}{r_{ij}^{3}} 
\cos\theta_i \sin \theta_j \sin\Delta\phi_{ij} 
\right)
 \\
 D_{\phi\phi}^{ii} 
 &=& \sum_{j(\ne i)}
 \left(
 \frac{3}{r_{ij}^{5}} 
 \left[\sin\theta_i \sin \theta_j \sin\Delta\phi_{ij} \right]^2
 - \frac{1}{r_{ij}^{3}} 
\sin\theta_i \sin \theta_j \cos\Delta\phi_{ij}
\right)
\\
 D_{\theta\theta}^{ij} 
 &=& 
 \frac{3}{r_{ij}^{5}} 
 \left[ - \cos\theta_i \sin \theta_j \cos\Delta\phi_{ij} + \sin\theta_i \cos\theta_j \right]
 \left[ - \sin\theta_i \cos \theta_j \cos\Delta\phi_{ij} + \cos\theta_i \sin\theta_j \right]
 \nonumber\\
& +& \frac{1}{r_{ij}^{3}} 
 \left[ \cos\theta_i \cos \theta_j \cos\Delta\phi_{ij} + \sin\theta_i \sin\theta_j \right]
\label{eq:Dijtt}
 \\
 D_{\theta\phi}^{ij} 
 &=&
 \frac{3}{r_{ij}^{5}} 
 \left[ - \cos\theta_i \sin \theta_j \cos\Delta\phi_{ij} + \sin\theta_i \cos\theta_j \right]
 \left[ - \sin\theta_i \sin \theta_j \sin\Delta\phi_{ij} \right]  
 + \frac{1}{r_{ij}^{3}} 
\cos\theta_i \sin \theta_j \sin\Delta\phi_{ij}
\label{eq:Dijtp}
 \\
 D_{\phi\phi}^{ij} 
 &=& 
 - \frac{3}{r_{ij}^{5}} 
 \left[\sin\theta_i \sin \theta_j \sin\Delta\phi_{ij} \right]^2
 + \frac{1}{r_{ij}^{3}} 
\sin\theta_i \sin \theta_j \cos\Delta\phi_{ij}
\label{eq:Dijpp}
\end{eqnarray}
\end{widetext}
with $i\ne j$ for Eqs.~(\ref{eq:Dijtt})-(\ref{eq:Dijpp}). Due to the rotational invariant around the $z$ axis, we can prove that $\sum_{j=1}^{N} D_{\theta\phi}^{ij} = \sum_{j=1}^{N} D_{\phi\phi}^{ij}=0$.

To determine the lowest energy structures for $10\le N\le 200$, we started various configurations of $(\theta_{i},\phi_{i})$ for $i=1, \cdots, N$, where $\theta_i$ and $\phi_i$ are random values restricted to $0<\theta_i<\pi$ and $0<\phi_{i}<2\pi$. More than 200 random distributions were considered to find the lowest energy structure for each $N$. The optimization of $V (\equiv E_{\rm opt})$ was performed by using the Broyden-Fletcher-Goldfarb-Shanno algorithm \cite{numerical_recipe}. We found that the lowest energy structures had no imaginary frequencies and the values of $E_{\rm opt}$s were exactly equal to those in the Cambridge Cluster Database (CCD) \cite{CCD} except for some $N$s. For $N=171, 177, 191$, and $197$, the $E_{\rm opt}$s obtained were higher than those in the CCD by less than 0.1: the difference was maximum at $N=191$, where $E_{\rm opt}=16783.5248378$ and $E_{\rm CCD}=16783.4522193$. We also found that for $N=177$ and 197, the value of $E_{\rm opt}$ is equal to that obtained by genetic algorithm approach \cite{morris}. These imply that the basin-hopping approach used in Ref.~\cite{wales} is suitable for finding the lowest energy structure in the Thomson problem. We thus referred to the CCD \cite{CCD} to obtain the position of charges for $N=171, 177, 191$, and 197. For a later use, we also referred to the CCD for $N=212+k, 252+k, 312+k, 272+k, 282+k$, and $372+k$ with $k=0,\pm 1$. When $\vert\sin\theta_i\vert < 10^{-6}$, we regarded that the charge $i$ is located at the north or south poles on a sphere and did not use Eq.~(\ref{eq:ddphi}) for $i$ in solving Eq.~(\ref{eq:eigen}) because $m_{i\phi}=0$. For each $N$, we obtained $\omega_{i}^{(N)}$ with $i=1, \cdots, 2N$ by solving Eq.~(\ref{eq:eigen}). The $\omega_{i}^{(N)}$s with $i=1,2$, and 3 are zero because the corresponding modes are the rotation around the $x, y$, and $z$ axes. Below, the maximum value of $\omega_{i}^{(N)}$ (i.e., $\omega_{2N}^{(N)}$) will be denoted as $\omega_{\rm max}(N)$ or simply $\omega_{\rm max}$.


\begin{figure}[tt]
\center
\includegraphics[scale=0.45]{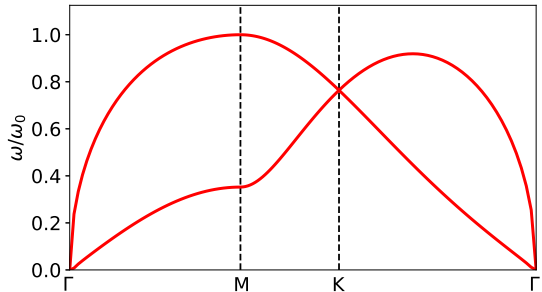}
\caption{The phonon dispersion curve of 2D triangular lattice. $\omega_0=Ca_{0}^{-3/2}$ with a unit mass and $C=3.383698$. } \label{fig_1} 
\end{figure}

\subsection{Lattice dynamics on 2D triangular lattice}
The optimized configuration on a unit sphere is like a triangular lattice, whereas deformation of a triangle will be observed due to a finite curvature, producing five-coordinated charges. For comparison, the lattice dynamics of the Coulomb crystal in the 2D triangular structure was considered. Due to the repulsive potential, the periodic boundary condition must be imposed to produce the dynamical stability of charged particles. The dynamical matrix was constructed by considering the central force potential model \cite{mermin}. For the wavevector $\bm{q}=(q_x,q_y)$, it is written as 
\begin{eqnarray}
 \tilde{D}_{\alpha\beta} (\bm{q}) = -2 \sum_{j\ne 0}\sin^2 \left(\frac{\bm{q}\cdot \bm{R}_j}{2}\right) 
 D_{\alpha\beta}^{j}
 \label{eq:2DHX}
\end{eqnarray}
with $\alpha,\beta=x,y$. The position of the $j$-th particle is given by $\bm{R}_j$, a linear combination of the primitive vectors $\bm{a}_1=(a_0,0)$ and $\bm{a}_2 = a_0(-1/2,\sqrt{3}/2)$ with the lattice constant $a_0$. The force constant matrix is expressed by
\begin{eqnarray}
 D_{\alpha\alpha}^{j} = \frac{1}{r_{j}^{3}} - \frac{3\alpha_{j}^{2}}{r_{j}^{5}}, \ \ 
 D_{\alpha\beta}^{j} = - \frac{3\alpha_{j}\beta_j}{r_{j}^{5}}
 \label{eq:FC}
\end{eqnarray}
with $r_j=\vert \bm{R}_j \vert$, $r_{j}^{2}=x_{j}^2+y_{j}^{2}$, and $\alpha\ne\beta$. In the present paper, no out-of-plane displacements were considered, whereas the 2D triangular lattice has shown to be unstable if the particle can move perpendicular to the 2D surface \cite{ono2021_LJ}. Due to the long-range Coulomb forces, the summation of Eq.~(\ref{eq:2DHX}) was taken over the particles up to $r_{j}= 2500a_0$. It should be noted that $a_0$ is only the characteristic length in the 2D triangular structure under the Coulomb potential. From Eqs.~(\ref{eq:2DHX}) and (\ref{eq:FC}), the phonon frequency $\omega$ is proportional to $a_{0}^{-3/2}$.

Figure \ref{fig_1} shows the dispersion curve of a 2D triangular lattice along the symmetry lines. At low $\omega$, the longitudinal and transverse branches show the different $q$ dependence: $\omega\propto q^{1/2}$ and $q$, respectively, with $q=\vert \bm{q}\vert$ \cite{bonsall}. The longitudinal modes have the maximum frequency $\omega_0$ at the point M$(0,1/3)$. 

\begin{figure}
\center
\includegraphics[scale=0.55]{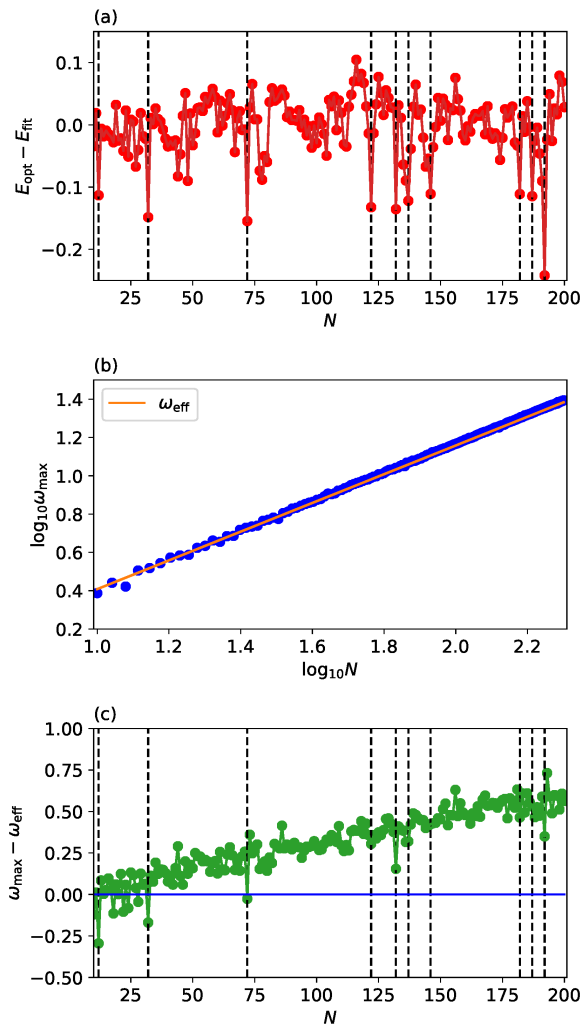}
\caption{The $N$-dependence of (a) $E_{\rm opt}-E_{\rm fit}$, (b) $\omega_{\rm max}$, and (c) $\omega_{\rm max}-\omega_{\rm eff}$, where $E_{\rm fit}$ and $\omega_{\rm eff}$ are expressed by Eqs.~(\ref{eq:energy_fit}) and (\ref{eq:fit}), respectively. Dashed lines in (a) and (c) indicate the magic numbers for $E_{\rm opt}$, i.e., $N=12, 32, 72, 122, 132, 137, 146, 182, 187$, and 192 identified in Ref.~\cite{morris}. } \label{fig_2} 
\end{figure}

\begin{figure}
\center
\includegraphics[scale=0.42]{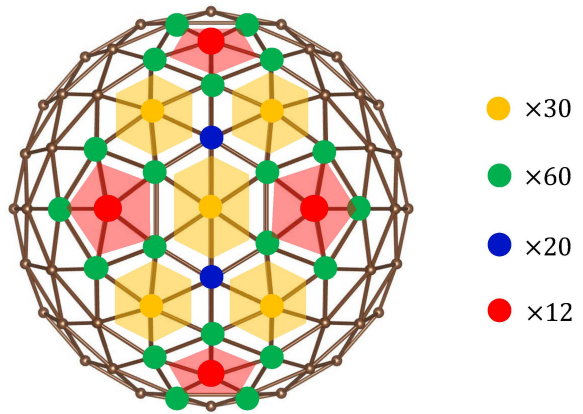}
\caption{The distribution of $N=122$ charges on a sphere. Four pentagons are colored red. Two pentagons share a hexagon colored orange. Three hexagons share a vertex colored blue. } \label{fig_3} 
\end{figure}

\section{Results and Discussion}
The optimized energy $E_{\rm opt}$ is proportional to $N^2/2$ that is exactly equal to the electrostatic energy stored by a spherical capacitor of a unit radius. The energy corrections proportional to $N^{3/2}$ and $N^{1/2}$ have been proposed in Refs.~\cite{morris,glasser},
\begin{eqnarray}
 E_{\rm fit}(N) = \frac{1}{2}\left(N^2 - a N^{3/2} + b N^{1/2}\right)
 \label{eq:energy_fit}
\end{eqnarray}
with the parameters of $a=1.10461$ and $b=0.137$. Figure \ref{fig_2}(a) shows the difference between $E_{\rm opt}$ and $E_{\rm fit}$ as a function of $N$. Anomalously low energies are found at $N=12, 32, 72, 122, 132, 137, 146, 182, 187$, and 192, which agree with the magic numbers identified in Ref.~\cite{morris}. These structures have icosahedral symmetry except for $N=137, 146, 182$, and $187$ \cite{CCD}. 

As shown in Fig.~\ref{fig_2}(b), the value of $\omega_{\rm max}$ also increases with $N$. To understand the $N$-dependence of $\omega_{\rm max}$, we consider the case that the averaged surface area $4\pi/N$ is equal to the unit cell area of the 2D triangular lattice $\sqrt{3}a_{\rm eff}^{2}/2$ with the effective lattice constant $a_{\rm eff}$. By performing the numerical calculations, we obtain the maximum frequency of 2D triangular lattice $C=3.383698$ when $a_0=1$ (see Fig.~\ref{fig_1}). We thus propose an effective frequency
\begin{eqnarray}
 \omega_{\rm eff} = Ca_{\rm eff}^{-3/2}, \ \ \ a_{\rm eff} = \left(\frac{8\pi}{\sqrt{3}N}\right)^{1/2},
 \label{eq:fit}
\end{eqnarray}
which behaves as $\omega_{\rm eff}\propto N^{3/4}$. The calculated $\omega_{\rm max}$ is well fitted by Eq.~(\ref{eq:fit}), indicating that the maximum frequency is approximately determined by the particle density. 

To find magic numbers, we plotted the difference between $\omega_{\rm max}$ and $\omega_{\rm eff}$ versus $N$ in Fig.~\ref{fig_2}(c). For large $N$, Eq.~(\ref{eq:fit}) underestimates $\omega_{\rm max}$, which will be due to the curvature effect: the maximum distance between charges on a sphere is twice a radius, whereas that on the 2D triangular lattice is infinite. A larger potential energy will be stored in the former system, giving rise to enhance the magnitude of force constants. More importantly, a significant decrease of $\omega_{\rm max}$ is observed at $N=12, 32, 72, 132$, and $192$ and these are also the magic numbers derived from the total energy calculations in Fig.~\ref{fig_2}(a). On the other hand, no significant decrease in $\omega_{\rm max}$ is found at $N=122, 137, 146, 182$, and $187$. The behavior at $N=122$ is anomalous because the icosahedral structure does not yield a small $\omega_{\rm max}$. 

To understand an origin of the magic numbers for $\omega_{\rm max}$, we focus on the distribution of $\varepsilon_i$ defined in Eq.~(\ref{eq:one}). Table \ref{table1} lists the $\varepsilon_i$s in an ascending order and its degeneracy for $N=12, 32, 72, 122, 132$, and $192$. For $N=12$, all the charges have the same energy because the charges are distributed at the vertices of a regular icosahedron. For larger $N$s, 12 charges have the highest $\varepsilon_i$ because they have not six but five coordination numbers arising from geometrical constraint. It is interesting that at $N=72, 132$, and $192$, the degeneracy for low $\varepsilon_s$s is 60 except for the highest $\varepsilon_i$. On the other hand, at $N=122$, the degeneracy is scattered, i.e., 20, 60, 30, and 12. This is visualized in Fig.~\ref{fig_3}, where the charges at the center of the hexagon colored orange are not equivalent to those colored blue. The one-particle energies are thus split into $\varepsilon_1$ and $\varepsilon_3$. In general, when strain stored is equally distributed over the structure, $\varepsilon_i$ will be strongly degenerated. We thus hypothesize that strongly degenerated $\varepsilon_i$ will cause small curvature of the PES around the equilibrium, yielding a small value of $\omega_{\rm max}$. In addition to the presence of a highly symmetric structure, strong degeneracies of $\varepsilon_{i}$s will be important in the appearance of magic numbers. 

To check the validity of our hypothesis, we next study the cases of $N=212+k, 272+k, 282+k$, and $372+k$ with $k=0,\pm 1$ because these $N$ charges form an icosahedral structure as the lowest energy state \cite{CCD}. We find that the values of $\varepsilon_i$s are strongly degenerated: the degeneracy is also 60 for low $\varepsilon_i$s and 12 for the highest $\varepsilon_i$ (see also Table \ref{table1}). The values of $\omega_{\rm max}$ for $k=0$ are smaller than those for $k=\pm 1$, as listed in Table \ref{table2}. We thus identify $N=212, 272, 282$, and $372$ as magic numbers. Note that a small value of degeneracy, such as 20 and 30 of $\varepsilon_1$ in $N=212$ and 282, respectively, listed in Table \ref{table1}, will have a minor contribution to $E_{\rm opt}$ and thus to $\omega_{\rm max}$ for $N\ge 200$.

For relatively large $N$s, it is nontrivial to find global minimum energy structures due to the presence of many local minima in the PES. For $400 \le N\le 4352$, the energetic stability has been investigated for only selected $N$s \cite{wales2009}. Only $N=1632$ and $1902$ have been identified to show an icosahedral structure as its lowest energy state, where topological defects different from a pentagonal shape shown in Fig.~\ref{fig_3} are located at 12 vertices of the icosahedral structure. Furthermore, as $N$ increases, grain boundaries start to appear in the Voronoi representation. It will be interesting to study how these defects influence the vibrational frequencies and/or the presence of magic numbers. 

\begin{table*}
\begin{center}
\caption{The values of $\varepsilon_i$ with $i=1$ to $7$ for several $N$s. The figure in parenthesis indicates the degeneracy.  }
{
\begin{tabular}{llllllll} \hline\hline
$N$ & $\varepsilon_1$ & $\varepsilon_2$ & $\varepsilon_3$ & $\varepsilon_4$ & $\varepsilon_5$ & $\varepsilon_6$ & $\varepsilon_7$ \\
\hline
12 \hspace{1mm} & 4.097 (12) \hspace{1mm} & - \hspace{1mm} & - \hspace{1mm} & - \hspace{1mm} & - \hspace{1mm} & - \hspace{1mm} & - \\
32 \hspace{1mm} & 12.872 (20) \hspace{1mm} & 12.901 (12) \hspace{1mm} & - \hspace{1mm} & - \hspace{1mm} & - \hspace{1mm} & - \hspace{1mm} & - \\
72 \hspace{1mm} & 31.307 (60) \hspace{1mm} & 31.379 (12) \hspace{1mm} & - \hspace{1mm} & - \hspace{1mm} & - \hspace{1mm} & - \hspace{1mm} & - \\
122 \hspace{1mm} & 54.854 (30) \hspace{1mm} & 54.903 (60) \hspace{1mm} & 54.912 (20) \hspace{1mm} & 55.029 (12) \hspace{1mm} & - \hspace{1mm} & - \hspace{1mm} & - \\
132 \hspace{1mm} & 59.630 (60) \hspace{1mm} & 59.665 (60) \hspace{1mm} & 59.777 (12) \hspace{1mm} & - \hspace{1mm} & - \hspace{1mm} & - \hspace{1mm} & - \\
192 \hspace{1mm} & 88.304 (60) \hspace{1mm} & 88.336 (60) \hspace{1mm} & 88.376 (60) \hspace{1mm} & 88.528 (12) \hspace{1mm} & - \hspace{1mm} & - \hspace{1mm} & - \\
212 \hspace{1mm} & 97.904 (20) \hspace{1mm} & 97.933 (60) \hspace{1mm} & 97.935 (60) \hspace{1mm} & 98.000 (60) \hspace{1mm} & 98.151 (12) \hspace{1mm} & - \hspace{1mm} & - \\
272 \hspace{1mm} & 126.836 (60) \hspace{1mm} & 126.847 (60) \hspace{1mm} & 126.886 (60) \hspace{1mm} & 126.904 (20) \hspace{1mm} & 126.954 (60) \hspace{1mm} & 127.146 (12) \hspace{1mm} & -\\ 
282 \hspace{1mm} & 131.672 (30) \hspace{1mm} & 131.689 (120) \hspace{1mm} & 131.724 (60) \hspace{1mm} & 131.787 (60) \hspace{1mm} & 131.976 (12) \hspace{1mm} & - \hspace{1mm} & - \\ 
372 \hspace{1mm} & 175.284 (60) \hspace{1mm} & 175.291 (60) \hspace{1mm} & 175.322 (60) \hspace{1mm} & 175.334 (60) \hspace{1mm} & 175.356 (60) \hspace{1mm} & 175.447 (60) \hspace{1mm} & 175.668 (12) \\ 
\hline\hline
\end{tabular}
}
\label{table1}
\end{center}
\end{table*}

\begin{table}
\begin{center}
\caption{The value of $\omega_{\rm max}(N)$ for $N=212+k, 272+k, 282+k$, and $372+k$ with $k=0,\pm 1$.}
{
\begin{tabular}{cccc} \hline\hline
$N$ \hspace{2mm} & $\omega_{\rm max}(N-1)$ \hspace{2mm} & $\omega_{\rm max}(N)$ \hspace{2mm} & $\omega_{\rm max}(N+1)$  \\
\hline
212 \hspace{1mm} & 25.903 \hspace{1mm} & 25.656 \hspace{1mm} & 26.066 \\ 
272 \hspace{1mm} & 31.209 \hspace{1mm} & 31.017 \hspace{1mm} & 31.258 \\ 
282 \hspace{1mm} & 32.019 \hspace{1mm} & 31.950 \hspace{1mm} & 33.540 \\ 
372 \hspace{1mm} & 39.858 \hspace{1mm} & 39.293 \hspace{1mm} & 39.535 \\ 
\hline\hline
\end{tabular}
}
\label{table2}
\end{center}
\end{table}

\begin{figure}[tt]
\center
\includegraphics[scale=0.5]{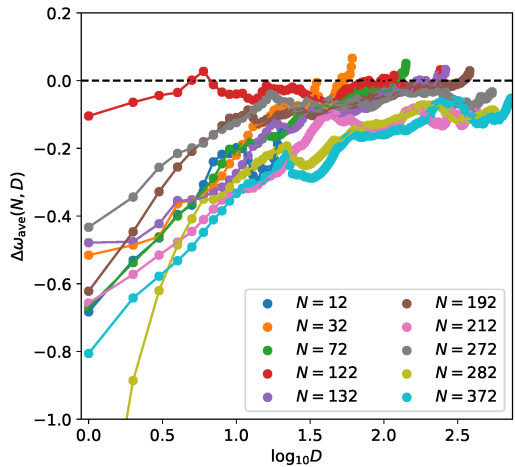}
\caption{The $D$-dependence of $\Delta\omega_{\rm ave}(N,D)$ for several $N$s. } \label{fig_4} 
\end{figure}

\begin{figure}[tt]
\center
\includegraphics[scale=0.5]{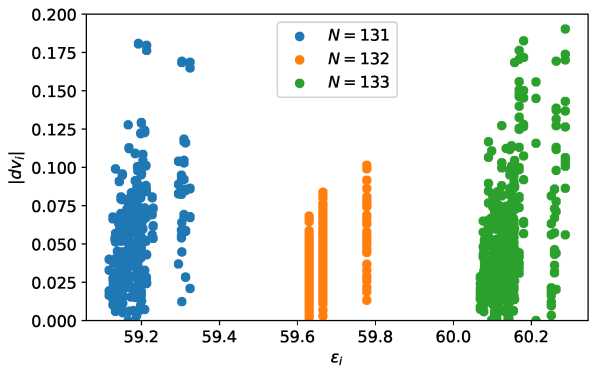}
\caption{The distribution of $\vert dv_{i}\vert$ versus $\varepsilon_i$ of the charge $i$ for $N=131, 132$, and 133. The data is plotted for the top four frequency modes. } \label{fig_5} 
\end{figure}

As a generalized problem, it will be valuable to investigate whether the magic number concept is valid when lower frequencies (i.e., $\omega_{i}^{(N)}$ with $i<2N$) are also considered. Interestingly, the maximum frequency tends to be highly degenerated at the magic numbers for vibrational frequency: for the top $D$ frequencies, the equalities $\omega_{2N}^{(N)}=\omega_{2N-1}^{(N)}=\cdots =\omega_{2N-D+1}^{(N)}$ can hold, where $D=2, 3, 4$, or 5. We first define an average of the top $D$ highest frequencies for $N$ charges as
\begin{eqnarray}
 \omega_{\rm ave} (N,D) = \frac{1}{D} \sum_{i=1}^{D} \omega_{2N+1-i}^{(N)}
\end{eqnarray}
with $D=1, \cdots, 2N$ and next define the difference as
\begin{eqnarray}
 \Delta \omega_{\rm ave} (N,D) = 2\omega_{\rm ave} (N,D) - \sum_{j=N\pm 1}\omega_{\rm ave} (j,D). 
\end{eqnarray}
If the magnitude of $\vert \Delta \omega_{\rm ave} (N,D)\vert$ is large, such an $N$ can be identified as a magic number. 

Figure \ref{fig_4} shows the $D$-dependence of $\Delta \omega_{\rm ave} (N,D)$ for several $N$s, where the value of $D$ is changed from 1 to $2(N-1)$ for each $N$. By assuming $D=2(N-1)\simeq 2N$, the value of $\Delta \omega_{\rm ave} (N,2(N-1))$ can be regarded as the difference between the average frequencies with an error of $\omega_{4}^{(N+1)}/(2N-2)\ll 1$. When $D=1$, the value of $\Delta \omega_{\rm ave} (N,D)$ is negatively large (less than $-0.4$) except for $N=122$, identifying the magic numbers as in Fig.~\ref{fig_2}(c) and Table \ref{table2}. When $D$ is increased, $\Delta \omega_{\rm ave} (N,D)$ approaches zero and may become positive for the largest $D$. In this way, for small $D$s, the magic number concept can hold, which may be due to the high degeneracy of the maximum frequency. On the other hand, for large $D$s, a small difference between $\omega_{\rm ave} (N,D)$s needs to be studied to identify the magic numbers. 

We finally remark on the oscillation amplitude of charges on a sphere. At a magic number of $N$ for vibrational frequency, the maximum frequency mode is a delocalized mode because the charge distribution is ordered compared to that of $N\pm 1$. This produces no anomalously large displacement of charges. To understand this, we consider the square of the displacement vector of the charge $i$ defined as
\begin{eqnarray}
 dv_{i}^{2} = (x_i-x_{i}^{0})^2 + (y_i-y_{i}^{0})^2 + (z_i-z_{i}^{0})^2,
\end{eqnarray}
where $(x_{i}^{0},y_{i}^{0},z_{i}^{0})$ and $(x_i,y_i,z_i)$ are the position of charges, respectively, before and after a displacement along the eigenvector, e.g., $z_{i}^{0}=\cos\theta_{i}^{0}$ and $z_{i}=\cos(\theta_{i}^{0}+u_{i\theta})$. As an example, we consider the cases of $N=132$ and $132\pm 1$, where the top four frequencies are degenerated at $N=132$. Figure \ref{fig_5} shows the distribution of $\vert dv_{i}\vert$ versus $\varepsilon_i$ for the top four frequencies ($4N$ points for each $N$). For $N=132$, the $\vert dv_{i}\vert$ of the charge $i$ having $\varepsilon_i$ is about less than 0.1. For $N=131$ and 133, the $\varepsilon_i$s are distributed around 59.2 and 60.1, respectively, and the distribution of $\vert dv_{i}\vert$ is elongated up to 0.2. It is important to find that only a few points have large $\vert dv_{i}\vert$. This implies that the non-magic number structures can have localized modes at high-frequency regimes.

\section{Conclusion}
We have studied the vibrational properties of $N$ charges on a sphere. The present paper shows that the $\omega_{\rm max}$ behaves as $N^{3/4}$, which is understood within the 2D triangular lattice model. The value of $\omega_{\rm max}$ is relatively small at $N=12, 32, 72, 132, 192, 212, 272, 282$, and $372$, so that these $N$s are identified as magic numbers. However, the total number of the magic numbers identified by the $\omega_{\rm max}$ calculations is smaller than that identified by the total energy calculations. This is because the value of $\omega_{\rm max}$ reflects a strong degeneracy of the one-particle energies and the icosahedral symmetry of distribution of $N$ charges. The presence of nonequivalent charges with the six-coordination number weakens the degeneracy, enhancing the value of $\omega_{\rm max}$ as in $N=122$, irrespective to the icosahedral structure. We have also demonstrated that the magic number concept can hold as long as an average of the top $D$ frequencies with a small $D$ is considered and that the charges at the magic numbers can have small oscillation amplitude for the maximum frequency modes.

The present paper will pave the way to future investigation concerning the magic numbers in realistic materials because the one-particle energy has been used in many model potentials such as Tersoff \cite{tersoff}, embedded-atom \cite{eam}, and neural-network potentials \cite{nn}. Also the present paper will provide another perspective to the Thomson problem from a point of view of magic numbers for vibrational frequency.



\begin{acknowledgments}
This study was supported by the a Grant-in-Aid for Scientific Research (C) (Grant No. 21K04628) from JSPS.
\end{acknowledgments}









\begin{thebibliography}{99}

\bibitem{romero} D. Romero, C. Barr\'{o}n, D. G\'{o}mez, The optimal geometry of Lennard-Jones clusters: 148-309, Comp. Phys. Comm. {\bf 123}, 87 (1999).

\bibitem{doye} J. P. K. Doye and L. Meyer, Mapping the magic numbers in binary Lennard-Jones clusters, Phys. Rev. Lett. {\bf 95}, 063401 (2005).

\bibitem{zhang} B. L. Zhang, C. H. Xu, C. Z. Wang, C. T. Chan, and K. M. Ho, Systematic study of structures and stabilities of fullerenes, Phys. Rev. B {\bf 46}, 7333 (1992). 

\bibitem{NP} F. Baletto and R. Ferrando, Structural properties of nanoclusters: Energetic, thermodynamic, and kinetic effects, Rev. Mod. Phys. {\bf 77}, 371 (2005).

\bibitem{wang} J. Wang, C. F. Mbah, T. Przybilla, B. A. Zubiri, E. Spiecker, M. Engel, and N. Vogel, Magic number colloidal clusters as minimum free energy structures, Nat. Commun. {\bf 9}, 5259 (2018).




\bibitem{smale} S. Smale, Mathematical problems for the next century, Math. Intell. {\bf 20}, 7 (1998).

\bibitem{morris} J. R. Morris, D. M. Deaven, and K. M. Ho, Genetic-algorithm energy minimization for point charges on a sphere, Phys. Rev. B {\bf 53}, R1740 (1996). 

\bibitem{alt} E. L. Altschuler, T. J. Williams, E. R. Ratner, R. Tipton, R. Stong, F. Dowla, and F. Wooten, Possible global minimum lattice configurations for Thomson's problem of charges on a sphere, Phys. Rev. Lett. {\bf 78}, 2681 (1997).

\bibitem{alt2} E. L. Altschuler and A. P. Garrido, Global minimum for Thomson's problem of charges on a sphere, Phys. Rev. E {\bf 71}, 047703 (2005).

\bibitem{wales} D. J. Wales and S. Ulker, Structure and dynamics of spherical crystals characterized for the Thomson problem, Phys. Rev. B {\bf 74}, 212101 (2006). 

\bibitem{wales2009} D. J. Wales and H. McKay, Defect motifs for spherical topologies, Phys. Rev. B {\bf 79}, 224115 (2009). 

\bibitem{mehta} D. Mehta, J. Chen, D. Z. Chen, H. Kusumaatmaja, and D. J. Wales, Kinetic Transition Networks for the Thomson Problem and Smale's Seventh Problem, Phys. Rev. Lett. {\bf 117}, 028301 (2016). 


\bibitem{peters1} V. A. Schweigret and F. M. Peeters, Spectral properties of classical two-dimensional clusters, Phys. Rev. B {\bf 51}, 7700 (1995). 

\bibitem{peters2} M. Kong, B. Partoens, A. Matulis, and F. M. Peeters, Structure and spectrum of two-dimensional clusters confined in a hard wall potential, Phys. Rev. E {\bf 69}, 036412 (2004). 

\bibitem{peters3} K. Nelissen, A. Matulis, B. Partoens, M. Kong, and F. M. Peeters, Spectrum of classical two-dimensional Coulomb clusters, Phys. Rev. E {\bf 73}, 016607 (2006). 

\bibitem{yang} W. Yang, M. Kong, M. V. Milo\ifmmode \check{s}\else \v{s}\fi{}evi\ifmmode \acute{c}\else \'{c}\fi{}, Z. Zeng, and F. M. Peeters, Two-dimensional binary clusters in a hard-wall trap: Structural and spectral properties, Phys. Rev. E {\bf 76}, 041404 (2007). 

\bibitem{robinson} M. Robinson, I. Suarez-Martinez, and N. A. Marks, Generalized method for constructing the atomic coordinates of nanotube caps, Phys. Rev. B {\bf 87}, 155430 (2013).

\bibitem{batle} J. Batle, O. Ciftja, M. Naseri, M. Ghoranneviss, A. Farouk, and M. Elhoseny, Equilibrium and uniform charge distribution of a classical two-dimensional system of point charges with hard-wall confinement, Phys. Scr. {\bf 92}, 055801 (2017).

\bibitem{yangQM} L. Yang and Z. Yao, Two and three electrons on a sphere: A generalized Thomson problem, Phys. Rev. B {\bf 97}, 235431 (2018). 

\bibitem{gnidovec} A. Gnidovec and C. \ifmmode \check{C}\else \v{C}\fi{}opar, Orientational ordering of point dipoles on a sphere, Phys. Rev. B {\bf 102}, 075416 (2020).


\bibitem{ziman} J. M. Ziman, {\it Electrons and Phonons} (Oxford University Press, New York, 1960).


\bibitem{numerical_recipe} W. M. Press, B. P. Flannery, S. A. Teukolsky, W. T. Vetterling, {\it Numerical Recipes in Fortran 90: The Art of Parallel Scientific Computing} (Cambridge University Press, Cambridge, 1996).

\bibitem{CCD} http://www-wales.ch.cam.ac.uk/CCD.html

\bibitem{mermin} N. W. Ashcroft, N. D. Mermin, and D. Wei, {\it Solid State Physics}, revised edition, (Cengage, Boston, 2016).

\bibitem{ono2021_LJ} S. Ono and T. Ito, Theory of dynamical stability for two- and three-dimensional Lennard-Jones crystals, Phys. Rev. B {\bf 103}, 075406 (2021).

\bibitem{bonsall} L. Bonsall and A. A. Maradudin, Some static and dynamical properties of a two-dimensional Wigner crystal, Phys. Rev. B {\bf 15}, 1959 (1977).

\bibitem{glasser} L. Glasser and A. G. Every, Energies and spacings of point charges on a sphere, J. Phys. A: Math. Gen. {\bf 25}, 2473 (1992). 


\bibitem{tersoff} J. Tersoff, New empirical approach for the structure and energy of covalent systems, Phys. Rev. B {\bf 37}, 6991 (1988).

\bibitem{eam} M. S. Daw and M. I. Baskes, Embedded-atom method: Derivation and application to impurities, surfaces, and other defects in metals, Phys. Rev. B {\bf 29}, 6443 (1984). 

\bibitem{nn} J. Behler and M. Parrinello, Generalized neural-network representation of high-dimensional potential-energy surfaces, Phys. Rev. Lett. {\bf 98}, 146401 (2007).


\end{thebibliography}
\end{document}